\documentclass[aps,preprint,prd,epsfig]{revtex4}
\usepackage{amssymb}
\usepackage{amsmath}
\usepackage{graphicx}
\usepackage{epsfig}
\usepackage{ccaption}
\usepackage{color}
\usepackage{fancyhdr}
\newcommand{\be}{\begin{equation}}
\newcommand{\ee}{\end{equation}}
\newcommand{\bea}{\begin{eqnarray}}
\newcommand{\eea}{\end{eqnarray}}
\newcommand{\bes}{\begin{subequations}}
\newcommand{\ees}{\end{subequations}}

\newcommand{\bc}{\begin{center}}
\newcommand{\ec}{\end{center}}
\begin{document}
\title{ Embedding  cosmological inflation, axion dark matter and seesaw mechanism  in a  3-3-1 gauge  model  }
\author{J. G. Ferreira, C. A de S. Pires, J. G. Rodrigues, P. S. Rodrigues da Silva}
\affiliation{{ Departamento de F\'{\i}sica, Universidade Federal da Para\'\i ba, Caixa Postal 5008, 58051-970,
Jo\~ao Pessoa, PB, Brazil}}

\date{\today}

\begin{abstract}
The  Peccei-Quinn symmetry  is an automatic symmetry of the 3-3-1 gauge models which, consequently are not plagued with the strong CP problem. Nevertheless,  the axion that emerges from spontaneous breaking of Peccei-Quinn symmetry cannot be made invisible in the original versions of these models, unless we extend their scalar sector by an additional neutral scalar singlet. In this case we show that  if, we also add  heavy neutrinos in the singlet form, we get to solve three interesting open questions at once: the real component of the  neutral scalar singlet  driving inflation, the axion playing the role of the dark matter of the universe and standard neutrinos gaining masses through  seesaw mechanism.
 \end{abstract}
\maketitle
\section{Introduction}
\label{secint}
The  $SU(3)_C \times SU(3)_L \times U(1)_N$ (3-3-1) gauge models for the electroweak interactions  are  interesting in their own right. For example, in these models generations cannot replicate unrestrictedly as in the standard model (SM), since they are not  exact replicas of one another and each is separately anomalous. However, when three generations are taken into account, gauge anomaly is automatically canceled~\cite{Frampton:1992wt}, providing a reason for the existence of three families of fermions. 

Also, the set  of constraints from gauge invariance of the Yukawa interactions together with those coming from the  anomaly cancellation conditions are enough to fix the electric charges  of the particles in the 3-3-1 model, thus providing an understanding of the pattern of electric charge quantization~\cite{deSousaPires:1998jc}\cite{ deSousaPires:1999ca}. 

In what concerns the  Peccei-Quinn~(PQ) symmetry, it is an automatic symmetry of these models,  thus elegantly solving the strong CP-problem~\cite{Pal:1994ba}. However, the original versions of the 3-3-1 gauge models furnish an unrealistic axion because of its sizable couplings with the standard particles~\cite{Weinberg:1977ma}\cite{Wilczek:1977pj}.  In order to we have an  invisible axion  a neutral scalar singlet must be added to the conventional scalar sector (three Higgs triplets)~\cite{Kim:1979if}\cite{Shifman:1979if}\cite{Dine:1981rt}.

Regarding neutrino masses,  canonical seesaw mechanisms, as  type I and type II, as well as the inverse seesaw mechanism are easily implemented in the framework of the 3-3-1 models~\cite{Montero:2001ts}\cite{Dias:2012xp}\cite{Dong:2008sw}\cite{Boucenna:2015zwa} . 
 
Last in the sequence but not least in importance, we remember that  some versions of the 3-3-1 models  have in their conventional particle content a stable and neutral particle that may play the role of cold dark matter in the WIMP form\cite{deS.Pires:2007gi}\cite{Mizukoshi:2010ky}\cite{daSilva:2014qba}\cite{Dong:2015rka}. These are only a few remarkable things that make 3-3-1 models interesting candidates for new physics beyond the SM. 

Although theoretical and phenomenological aspects of the 3-3-1 gauge models have received considerable attention, cosmological issues have been much less developed, in particular, the inflationary scenario was scarcely studied in a supersymmetric version of 3-3-1 model~\cite{Huong:2008ia,Long:2015qza}.
Thus, it would be very interesting if inflation could be implemented in the framework of other 3-3-1 gauge  models, even without supersymmetry.

At this point we would like to  remark that a possible way of providing a common origin to cosmological inflation,  the cold dark mater, neutrino masses and  solution to the strong CP-problem is by adding exotic vector like quarks, right-handed neutrinos and  neutral scalar singlet to the standard model, building a scenario called SMASH~\cite{Dias:2014osa}\cite{Barenboim:2015cqa}\cite{Ballesteros:2016euj}\cite{Ballesteros:2016xej}. This  packet of new  particles together with adequate interactions allows the  implementation of the PQ symmetry in  the standard models. The PQ symmetry is spontaneously broken when the neutral scalar singlet develop vacuum expectation value (VEV) different from zero. In this circumstance, the imaginary part of the neutral scalar singlet  will be the invisible axion, which may play the role of dark matter, while the real part may play the role of the inflaton. Moreover, on coupling  the neutral scalar to the right-handed neutrino, through an Yukawa interaction, the VEV of the neutral scalar, that is around $(10^{10}  -10^{11})$~GeV, will generate  heavy neutrinos that may trigger the type I seesaw mechanism yielding small neutrino masses for the standard neutrinos. The problem with this scenario is that it generates an inflaton potential of the type $\lambda \phi^4$ which is practically excluded by the current bounds from PLANCK15~\cite{Ade:2015lrj}. A way of circumventing such a problem is considering that the inflaton couples non-minimally with  the scalar curvature $R$ or taking into account radiative corrections to the inflaton  potential. The question we follow here is that if it is possible to implement such a scenario into the 3-3-1 model framework, since PQ symmetry is an automatic symmetry of the model.
In this paper we show that this  is possible when  radiative corrections are taken into account  for the inflaton potential. 

The paper is divided in the following way: In  Sec.~\ref{sec1} we revisit the 3-3-1 model that contains an invisible axion in its spectrum. Next, in Sec.~\ref{sec2}, we develop the inflationary paradigm in such model.  We finally conclude in Sec.~\ref{sec3}.
\section{The 3-3-1 model, the Pecei-Quinn symmetry and the invisible axion }
\label{sec1}
The  model developed here is  one  proposed in Ref.~\cite{Dias:2003iq} which is a modification of the original one~\cite{Singer:1980sw}\cite{Montero:1992jk}\cite{Foot:1994ym}.  The first modification  is  in the leptonic sector where heavy right-handed neutrinos in the singlet form are added to the model,
\bea f^a_L = \left (
\begin{array}{c}
\nu^a_L \\
e^a_L \\
(\nu^{c}_R)^a
\end{array}
\right )\sim(1\,,\,3\,,\,-1/3)\,,\,\,\,e_{aR}\,\sim(1,1,-1) \,,\,\,\,N_{aR}\,\sim(1,1,0) \eea
with $a=1,2,3$ representing the three known generations. We are
indicating the transformation under 3-3-1 after the similarity
sign, ``$\sim$''. 

The  quark sector is kept intact with  one generation of left-handed fields coming in
the triplet fundamental representation of $SU(3)_L$ and the other
two composing an anti-triplet  representation with the content,
\bea &&Q_{iL} = \left (
\begin{array}{c}
d_{iL} \\
-u_{iL} \\
d^{\prime}_{iL}
\end{array}
\right )\sim(3\,,\,\bar{3}\,,\,0)\,,\,\,\,Q_{3L} = \left (
\begin{array}{c}
u_{3L} \\
d_{3L} \\
u^{\prime}_{3L}
\end{array}
\right )\sim(3\,,\,3\,,\,1/3)\,, \label{quarks1} \eea
and the right-handed fields,
\bea
&&u_{iR}\,\sim(3,1,2/3)\,,\,d_{iR}\,\sim(3,1,-1/3)\,,\, d^{\prime}_{iR}\,\sim(3,1,-1/3)\nonumber \\
&&u_{3R}\,\sim(3,1,2/3)\,,\,d_{3R}\,\sim(3,1,-1/3)\,,\,u^{\prime}_{3R}\,\sim(3,1,2/3),
\label{quarks2} \eea
where $j=1,2$ represent different generations. The primed quarks
are the exotic ones but with the usual electric charges.

In order to generate the masses for the gauge bosons and fermions,
the model requires only  three Higgs scalar triplets. For our proposal here we add a neutral scalar singlet to these triplets transforming in the following way by the 3-3-1 symmetry, 
\bea 
&&\chi = \left (
\begin{array}{c}
\chi^0 \\
\chi^{-} \\
\chi^{\prime 0}
\end{array}
\right )\sim(1\,,\,3\,,\,-1/3) \,\,,\,\, \eta = \left (
\begin{array}{c}
\eta^0 \\
\eta^- \\
\eta^{\prime 0}
\end{array}
\right )\sim(1\,,\,3\,,\,-1/3),\nonumber \\
&& \rho = \left (
\begin{array}{c}
\rho^+ \\
\rho^0 \\
\rho^{\prime +}
\end{array}
\right )\sim(1\,,\,3\,,\,2/3)\,,\,\,\,\,\,\,\,\phi\sim(1,1,0). \label{chieta}
\eea
Thus the particle content of the model is extended by the fields $N_{ a_R}$  and $\phi$ .  
 
In order to keep intact the physics results of the Ref.~\cite{Dias:2003iq}, the  Lagrangian of the model  must be invariant by the following set of discrete symmetries $Z_{11} \otimes Z_2$ but now with  $Z_{11}$ acting as
\bea \phi\,\,\,\,\,&\rightarrow &\,\,\,\,\,
\omega^{-1}_1\phi\,,\,\,\,\,\,\,\,\,\,\,\,\,\,\,\,\,\,\,\,\,\,\,\,\,\,f_{aL}\,\,\,\,\,\,\,
\rightarrow\,\,\,\,\,\,\,\omega_1f_{aL}\,,
\nonumber \\
\rho\,\,\,\,\,&\rightarrow &\,\,\,\,\,
\omega^{-1}_2\rho\,,\,\,\,\,\,\,\,\,\,\,\,\,\,\,\,\,\,\,\,\,\,\,\,\,\,d_{aR}\,\,\,\,\,\,\,
\rightarrow\,\,\,\,\,\,\,\omega_2 d_{aR}\,,
\nonumber \\
\chi\,\,\,\,\,&\rightarrow &\,\,\,\,\,
\omega^{-3}_3\chi\,,\,\,\,\,\,\,\,\,\,\,\,\,\,\,\,\,\,\,\,\,\,\,u_{3R}^\prime
\,\,\,\,\,\,\,\,\rightarrow\,\,\,\,\,\,\,\omega_3u_{3R}^\prime\,,
\nonumber \\
Q_{iL}\,\,\,\,\,&\rightarrow &\,\,\,\,\, \omega^{-1}_4
Q_{iL}\,,\,\,\,\,\,\,\,\,\,\,\,\,\,\,\,\,\,\,\,
d_{iR}^\prime\,\,\,\,\,\,\,\rightarrow\,\,\,\,\,\,\,\omega_4d_{iR}^\prime\,,
\nonumber \\
\eta\,\,\,\,\,&\rightarrow &\,\,\,\,\,
\omega^{-1}_5\eta\,,\,\,\,\,\,\,\,\,\,\,\,\,\,\,\,\,\,\,\,\,\,\,\,\,\,u_{aR}\,\,\,\,\,\,\,
\rightarrow\,\,\,\,\,\,\,\omega_5u_{aR}\,,
\nonumber \\
Q_{3L}\,\,\,\,\,&\rightarrow &\,\,\,\,\, \omega_0 Q_{3L}\,,\,\,\,\,\,\,\,\,\,\,\,\,\,\,\,\,\,\,\, N_R\,\,\,\,\,\rightarrow \,\,\,\,\,
\omega^{-1}_5 N_R\,, \nonumber \\
e_{aR}\,\,\,\,\,\,\,
&\rightarrow&\,\,\,\,\,\,\,\omega_3e_{aR}\,,
\label{z11cargas}
 \eea
where 
$\omega_k \equiv e^{2\pi i\frac{k}{11}}\,,\,\{k=0,\pm 1,..., \pm
5\}$.

The  $Z_2$ symmetry  must act as
\be (\rho\,,\chi\,,d_R^\prime\,,u_{3R}^\prime\,,u_R\,,d_R\,,e_R )
\,\,\,\,\rightarrow\,\,\,\,
-(\rho\,,\chi\,,d_R^\prime\,,u_{3R}^\prime\,,u_R\,,d_R\,,e_R ) \,. \label{z2} \ee
These discrete  symmetries yield  the following Yukawa couplings
\bea {\cal L}^Y&&= G_1\bar{Q}_{3L}u^{\prime}_{3R} \chi
+G_2^{ij}\bar{Q}_{iL}d^{\prime}_{jR}\chi^* +
G_3^{3a}\bar{Q}_{3L}u_{aR}\eta +G_4^{ia}\bar{Q}_{iL}d_{aR}\eta^*
\nonumber \\
&&+G_5^{3a}\bar{Q}_{3L}d_{aR}\rho
+G_6^{ia}\bar{Q}_{iL}u_{aR}\rho^*
+g_{ab}\bar{f}_{aL}e_{bR}\rho + h_{ab}\bar{f}_{aL}\eta N_{bR} +h^{\prime}_{ab} \phi \bar{N}^C_{aR}N_{bR} 
+\mbox{H.c.}. \label{yukintera} \eea
The transformations displayed in Eqs.~(\ref{z11cargas}) and (\ref{z2}) are a little different from the original case~\cite{Dias:2003iq}. The reason  of the modification is to accommodate the last two  terms in the Lagrangian above. These terms are crucial for our proposal, as we will see later.

The allowed renormalizable and gauge
invariant potential for this model is exactly the same as in the original case, i.e,
\bea V_H &=& \mu_\phi^2 \phi^2 + \mu_\chi^2 \chi^2
+\mu_\eta^2\eta^2 +\mu_\rho^2\rho^2+\lambda_1\chi^4
+\lambda_2\eta^4
+\lambda_3\rho^4+
\lambda_4(\chi^{\dagger}\chi)(\eta^{\dagger}\eta)
+\lambda_5(\chi^{\dagger}\chi)(\rho^{\dagger}\rho) \nonumber \\
&&+\lambda_6
(\eta^{\dagger}\eta)(\rho^{\dagger}\rho)+
\lambda_7(\chi^{\dagger}\eta)(\eta^{\dagger}\chi)
+\lambda_8(\chi^{\dagger}\rho)(\rho^{\dagger}\chi)+\lambda_9
(\eta^{\dagger}\rho)(\rho^{\dagger}\eta)+
\lambda_{10} (\phi\phi^*)^2
\nonumber \\
&& +\lambda_{11}(\phi\phi^*)(\chi^{\dagger}\chi)
+\lambda_{12}(\phi\phi^*)(\rho^{\dagger}\rho) +
\lambda_{13}(\phi\phi^*)(\eta^{\dagger}\eta)+
\lambda_{\phi}\epsilon^{ijk}\eta_i\rho_j\chi_k \phi+ H.c\,, 
\label{VH} \eea

Other tiny change arises in the definition of the PQ charges. In order to have chiral quarks under
$U_{PQ}(1)$, we need the following transformation
\bea u_{aL}&\rightarrow & e^{-i\alpha X_u}u_{aL}\,,\,\,\,\,
u_{aR}\,\,\,\rightarrow\,\,\, e^{i\alpha X_u}u_{aR}\,, \nonumber \\
u_{3L}^\prime &\rightarrow & e^{-i\alpha
X_u^\prime}u_{3L}^\prime\,,\,\,\,\,
u_{3R}^\prime\,\,\,\rightarrow\,\,\,
e^{i\alpha X_u^\prime}u_{3R}^\prime\,, \nonumber \\
d_{aL}&\rightarrow & e^{-i\alpha X_d}d_{aL}\,,\,\,\,\,
d_{aR}\,\,\,\rightarrow\,\,\, e^{i\alpha X_d}d_{aR}\,, \nonumber \\
d_{iL}^\prime &\rightarrow & e^{-i\alpha
X_d^\prime}d_{iL}^\prime\,,\,\,\,\,\,\,
d_{iR}^\prime\,\,\,\rightarrow\,\,\, e^{i\alpha
X_d^\prime}d_{iR}^\prime\,. \label{pqcargasQ} \eea

For the leptons we can define their PQ charges by,
\bea e_{aL}&\rightarrow & e^{i\alpha X_e}e_{aL}\,,\,\,\,\,
e_{aR}\,\,\,\rightarrow\,\,\, e^{i\alpha X_{eR}}e_{aR}\,,\,\,\, N_{aL} \rightarrow  e^{i\alpha X_N}N_{aL} \nonumber \\
\nu_{aL} &\rightarrow & e^{i\alpha X_\nu}\nu_{aL}\,,\,\,\,\,
\nu_{aR}\,\,\,\rightarrow\,\,\, e^{i\alpha X_{\nu R}}\nu_{aR}\,.
\label{pqcargasL} \eea

With these assignments and taking the Yukawa interactions in
Eq.~(\ref{yukintera}) into account, as well as the non-hermitean
terms $\eta\rho\chi\phi$, we easily see that the PQ charges for
the scalars are constrained and imply the following relations:
\be
X_d=-X_u\,,\,\,\,\,X_{d^\prime}=-X_{u^\prime}\,,\,\,\,\,X_\nu=X_{eR}
\,,\,\,\,\,X_e=X_{\nu R}\,. \label{pqvinculos} \ee
We can make the further choice $X_d=X_{d^\prime}$, leading to
\be X_d=X_{d^\prime}=-X_u=-X_{u^\prime}=-X_e=X_{eR}=X_\nu=-X_{\nu
R}=X_{N}\,, \label{pqc} \ee
implying that the PQ symmetry is chiral for the leptons too. The scalars transform as
\bea \phi &\rightarrow & e^{-2i\alpha
X_d}\phi\,,\,\,\,\,\,\,\,\,\,\,\,\,\,\,
\eta^0\,\,\,\,\rightarrow\,\,\, e^{2i\alpha X_d}\eta^0 \nonumber \\
\eta^-&\rightarrow &
\eta^-\,,\,\,\,\,\,\,\,\,\,\,\,\,\,\,\,\,\,\,\,\,\,\,\,\,\,\,\,\,\,
\eta^{\prime 0}\,\,\,\rightarrow\,\,\, e^{2i\alpha X_d}\eta^{\prime 0} \nonumber \\
\rho^+&\rightarrow &
\rho^+\,,\,\,\,\,\,\,\,\,\,\,\,\,\,\,\,\,\,\,\,\,\,\,\,\,\,\,\,\,\,\,
\rho^{0}\,\,\,\,\rightarrow\,\,\, e^{-2i\alpha X_d}\rho^{0} \nonumber \\
\rho^{\prime +}&\rightarrow & \rho^{\prime
+}\,,\,\,\,\,\,\,\,\,\,\,\,\,\,\,\,\,\,\,\,\,\,\,\,\,\,\,\,\,\,
\chi^{0}\,\,\,\rightarrow\,\,\, e^{2i\alpha X_d}\chi^{0} \nonumber \\
\chi^{-}&\rightarrow &
\chi^{-}\,,\,\,\,\,\,\,\,\,\,\,\,\,\,\,\,\,\,\,\,\,\,\,\,\,\,\,\,\,\,
\chi^{\prime 0}\,\,\,\rightarrow\,\,\, e^{2i\alpha X_d} \chi^{\prime 0} \nonumber .\\ 
\label{pqcargasS} \eea

It is now clear that the entire Lagrangian of the model is
$U_{PQ}(1)$ invariant, providing a natural solution to the strong-CP problem.

To accomplish our proposal, let us consider that only $
\chi^{\prime 0} $, $ \rho^0 $, $\eta^0$ and $\phi$ develop  VEV and expand such fields around them
in the standard way,
\bea \chi^{\prime 0} &=&  \frac{1}{\sqrt{2}} (v_{\chi^{\prime}}
+R_{\chi^{\prime}} +iI_{\chi^{\prime}})\,,\,\,\,\,\,\,\, \eta^ 0=
\frac{1}{\sqrt{2}} (v_{\eta} +R_{\eta} +iI_{\eta})\,, \nonumber \\
\rho^0 &=& \frac{1}{\sqrt{2}} (v_{\rho} +R_{\rho}
+iI_{\rho})\,,\,\,\,\,\,\,\,\,\,\,\,\,\,\,\, \phi=
\frac{1}{\sqrt{2}} (v_{\phi} +R_{\phi} +iI_{\phi})\,.
\label{vacua} \eea
With such expansion, we obtain the set of constraint equations that guarantee that the potential has a minimum
\bea &&\mu^2_\chi +\lambda_1 v^2_{\chi^{\prime}} +
\frac{\lambda_4}{2}v^2_\eta +
\frac{\lambda_5}{2}v^2_\rho+\frac{\lambda_{11}}{2}v_\phi^2
+\frac{A}{v_{\chi^\prime}^2} =0,\nonumber \\
&&\mu^2_\eta +\lambda_2v^2_\eta + \frac{\lambda_4}{2}
v^2_{\chi^{\prime}} +\frac{\lambda_6}{2}v^2_\rho
+\frac{\lambda_{13}}{2} v^2_{\phi}+\frac{A}{v_\eta^2} =0,\nonumber \\
&&\mu^2_\rho +\lambda_3 v^2_\rho + \frac{\lambda_5}{2}
v^2_{\chi^{\prime}}
+\frac{\lambda_6}{2}v^2_\eta+\frac{\lambda_{12}}{2}
v^2_{\phi}+\frac{A}{v_\rho^2} =0,
\nonumber \\
&&\mu^2_\phi +\lambda_{10}v^2_{\phi} +
\frac{\lambda_{11}}{2}v^2_{\chi^{\prime}}+\frac{\lambda_{12}}{2}v^2_\rho+
\frac{\lambda_{13}}{2}v^2_\eta +\frac{A}{v_\phi^2}
=0\,,\label{mincondI} \eea
where we have defined $A\equiv\lambda_{\phi}v_\eta v_\rho
v_{\chi^{\prime}}v_\phi$.  The physical scalars are obtained by substituting these constraints into the mass matrices given by the second derivative of the potential. The axion arises from the mass matrix
$M_I^2$ given by,
\be - \frac{A}{2}\left(\begin{array}{cccc}  \frac{1}{
v_{\chi^\prime}^2} & \frac{1}{ v_\eta v_{\chi^\prime}} &
\frac{1}{ v_\rho v_{\chi^\prime}}
& \frac{1}{ v_{\chi^\prime}v_\phi} \\
\frac{1}{ v_\eta v_{\chi^\prime}} &  \frac{1}{ v_\eta^2} &
\frac{1}{ v_\eta v_\rho} & \frac{1}{ v_\eta v_\phi}\\
\frac{1}{ v_\rho v_{\chi^\prime}} & \frac{1}{ v_\eta v_\rho}
& \frac{1}{ v_\rho^2} & \frac{1}{ v_\eta v_\rho} \\
\frac{1}{ v_{\chi^\prime}v_\phi} & \frac{1}{ v_\eta v_\phi} &
\frac{1}{ v_\eta v_\rho} & \frac{1}{ v_\phi^2}
\end{array}
\right)\, \label{matrixI2} \ee
in the basis $(I_{\chi^\prime}\,,I_\eta\,,I_\rho\,,I_\phi )$. Its diagonalization furnishes an axion given by,  $a =
\frac{1}{\sqrt{1+\frac{v_{\chi^\prime}^2}{v_\phi^2}}}\left(I_\phi
-\frac{v_{\chi^\prime}}{v_\phi}I_{\chi^\prime}\right)$. As $v_\phi \gg v_{\chi^\prime}$ we have that $a \sim I_\phi$.

Now let us focus on the CP-even component of $\phi$. It will be our inflaton candidate. It composes the following mass matrix $M_R^2$ given by
\bea
\left(\begin{array}{cccc}  2\lambda_1
v^2_{\chi^{\prime}}-\frac{A}{2 v_{\chi^\prime}^2} &
\frac{\lambda_4 v_{\chi^{\prime}}v_\eta}{2}+\frac{A}{2 v_\eta
v_{\chi^\prime}} & \frac{\lambda_5 v_{\chi^{\prime}}v_\rho}{2}
+\frac{A}{2 v_\rho v_{\chi^\prime}} & \frac{A}{2 v_\phi v_{\chi^\prime}} \\
\frac{\lambda_4 v_{\chi^{\prime}}v_\eta}{2}+\frac{A}{2 v_\eta
v_{\chi^\prime}} & 2\lambda_2 v^2_\eta -\frac{A}{2 v_\eta^2} &
\frac{\lambda_6 v_\eta
v_\rho}{2} +\frac{A}{2 v_\rho v_\eta} & \frac{A}{2 v_\eta v_\phi}  \\
\frac{\lambda_5 v_{\chi^{\prime}}v_\rho}{2} +\frac{A}{2 v_\rho
v_{\chi^\prime}} & \frac{\lambda_6 v_\eta v_\rho}{2} +\frac{A}{2
v_\rho v_\eta} &2 \lambda_3v^2_\rho-\frac{A}{2 v_\rho^2} &
\frac{A}{2 v_\rho v_\phi} \\
\frac{A}{2 v_\phi v_{\chi^\prime}} & \frac{A}{2 v_\eta v_\phi} &
\frac{A}{2 v_\rho v_\phi} & 2\lambda_{10} v_\phi^2 -\frac{A}{2
v_\phi^2}
\end{array}
\right) \label{matrixR2}
\eea
in the basis $(R_{\chi^{\prime}}\,,R_\eta\,,R_\rho\,,R_\phi )$. As $v_\phi \gg v_\rho \,,\, v_\eta \,,\, v_{\chi^{\prime}}$, we have that  $R_\phi$ decouples  and its mass  is predicted to be $ m^2_{R_\phi}\sim2\lambda_{10}v^2_\phi$.

We would like to stress that  all the previous results  of Ref.~\cite{Dias:2003iq} concerning the solution to the strong CP-problem and to the axion profile remain valid here, namely, for an acceptable solution to the strong CP-problem, the $Z_{11}$ discrete symmetry implies  $ \theta_{eff}<10^{-9}$ which translates into  $v_\theta \leq 10^{10}$~GeV. The  axion  is  the imaginary component of the neutral scalar singlet $\phi$. It is invisible and free of domain wall problems.  The presence of large discrete symmetries stabilizes the axion against quantum gravity effects. At this point it is important to remark that the incorporation of PQ symmetry in 3-3-1 model as done here has as main purpose the explanation of the strong CP-problem and, as a byproduct, the fact that  the invisible axion fulfill the conditions to be a  viable dark matter candidate.

\section{Implementing Inflation }
\label{sec2}
Here we consider inflation in the specific framework of the 3-3-1 model presented in the previous section. Our aim is to  show that the real component of the $\phi$ field  will play the role of the inflaton with its potential satisfying  the slow roll conditions while providing the current prediction for the scalar spectral index, $n_s$,  and obeying the current bound on the  scalar to tensor ratio, $r$.

First thing to note is that the $\phi$ potential involves  the  terms,
\begin{eqnarray}
V_\phi &=& \mu_\phi^2 \phi^2 +
\lambda_{10} (\phi\phi^*)^2
+\lambda_{11}(\phi\phi^*)(\chi^{\dagger}\chi) \nonumber \\
&& +
\lambda_{12}(\phi\phi^*)(\rho^{\dagger}\rho) +
\lambda_{13}(\phi\phi^*)(\eta^{\dagger}\eta)+
\lambda_{\phi}\epsilon^{ijk}\eta_i\rho_j\chi_k \phi+ H.c. 
\label{vphi1}
\end{eqnarray}
However, as  $v_\phi >> v_{\eta\,,\,\rho\,,\,\chi}$, the terms in this  potential  that really  matter during inflation are
\begin{equation}
V_\phi = \mu_\phi^2 \phi^2 +
\lambda_{10} \phi^4. 
\label{vphi2}
\end{equation}
This is  the well known chaotic inflation scenario with the inflaton being the real part of $\phi$. From now on we use the notation $R_\phi \equiv \Phi$

A VEV around   $10^{10}$~GeV for $\phi$   implies that the dominant term in the above potential  is  $\lambda_{10}\phi^4$. However, as we know, the $\lambda_{10}\phi^4$ chaotic inflation is not favored by recent values of $r$ measured by PLANCK2015~\cite{Ade:2015lrj}. Thus, in order to circumvent this problem, we take into account radiative corrections to the potential which now reads, 
\begin{eqnarray}
 V(\Phi)=V_{tree} + V_{eff},
\label{realpotential}
\end{eqnarray}
with $V_{tree}=\lambda_{10}\Phi^4$ and $V_{eff}$  being the radiative corrections due to the coupling of $\Phi$ to the particle content of the 3-3-1 model.  The radiative corrections are engendered by the couplings of our inflaton with the right-handed neutrinos and the   scalars whose intensity is determined by the parameters $\lambda_{11}\,,\,\lambda_{12}\,,\,\lambda_{13}\,,\,\lambda_{\phi}$ and   $h^{\prime}$. As we will see below, reheating implies $\lambda_{\phi \,, 11\,,\,12\,,\,13}$ be very small. Thus the intensity of the radiative corrections is practically determined by $h^{\prime}$ which is the   coupling of the inflaton, $\Phi$,  to the heavy neutrino, $N_R$, and is given by the last term of the Lagrangian in Eq. (\ref{yukintera}).  As it is usual, here we  follow the  approach of Coleman and Weinberg whose expression to $V_{eff}$ is given by \cite{Coleman:1973jx} 
\begin{equation}
 V_{eff} = \frac{1}{64\pi^2} \sum_{i}\left[ (-1)^{2J}(2J+1)m^4_i\ln{\frac{m^2_i}{\Delta^2}}\right],
\end{equation}
where $m_i$ is the $\phi$-field dependent mass where $i=\eta\,,\,\rho\,,\,\chi\,,\,\phi\,,\,N_R$.  $J$ is  the spin of the respective contribution.  In our case   $N_R$  gives the dominant contribution, which is the only one we have to consider and amounts to take $ m_{N_R}=-\sqrt{2}h^{\prime} \Phi$. In this circumstance,  for our proposal here it is just sufficient to consider one family of heavy neutrinos.  After all this the potential that really matters during the inflationary period is given by,
\begin{equation}
 V(\Phi) \approx \lambda^{\prime}\left(\Phi^{4} + a^{\prime}\Phi^4 \ln{\frac{\Phi}{\Delta}}\right), \label{potaprox}
\end{equation}
where $\lambda^{\prime}=\frac{\lambda_{10}}{4}$ and  $a^{\prime}= \frac{a+160 \lambda^{\prime 2}}{32\pi^2\lambda^{\prime}} \approx \frac{a}{32\pi^2\lambda^{\prime}}$.  $\Delta$ is a renormalization scale. This approximation is justified because the amplitude of curvature perturbation demands a small $\lambda_{10}$.
The term $a$ carries the radiative contribution and in our case it is given by 
$a=-16 h^{\prime 4}$.  The negative sign is a characteristic feature of fermion contributions. Throughout this section we follow the approach given in Refs. \cite{NeferSenoguz:2008nn}\cite{Boucenna:2014uma}

\par We can now treat the issue of inflation, which occurs as long as the slow roll approximation is satisfied ($\epsilon \ll 1$, 
$\eta \ll 1$, $\zeta^2 \ll 1$). The slow roll parameters are given by~\cite{Liddle:1994dx}
\begin{eqnarray}
 &\epsilon \left(\phi_{R}\right)& = \frac{m^2_{P}}{16\pi}\left(\frac{ V^{\prime}}{ V}\right)^2, \quad \quad
 \eta \left(\phi_{R}\right) = \frac{m^2_{P}}{8\pi}\left(\frac{ V^{\prime \prime}}{ V}\right), \nonumber \\
 &\zeta^2  \left(\phi_{R}\right)& = \frac{m^4_{P}}{64\pi^2}\left(\frac{ V^{\prime \prime \prime}V^{\prime \prime}}{ V^2}\right),
\end{eqnarray}
where $m_P=1.22 \times 10^{19}$~GeV. 

The spectral index $n_S$, 
the scalar to tensor  ratio $r$ and the running of spectral index $\alpha \equiv \frac{dn_S}{d\ln{k}}$ are defined as~\cite{Lyth:2009zz}
\begin{eqnarray}
 &n_S&=1-6\epsilon+2\eta, \quad \quad \quad r=16\epsilon, \nonumber \\
 &\alpha&=16\epsilon \eta -24\epsilon^2-2\zeta^2.
\end{eqnarray}

For a wave number $k=0.05$  $Mpc^{-1}$, the Planck results indicate $n_S=0.9644 \pm 0.0049$ and $r<0.149$~\cite{Ade:2015lrj}.

\par The number of e-folds is given by
\begin{equation}
N = \frac{-8\pi}{m^2_P}\int_{\Phi_i}^{\Phi_f}\frac{V}{V^{\prime}}d\Phi,\label{efold}
\end{equation}
where $\Phi_{f}$ marks the end of inflation and is defined by $(\epsilon,\eta,\zeta^2)=1$. 
To find $\Phi_i$ we set  $N=50, 60$ and $70$ and solve Eq.~(\ref{efold})  for $\Phi_i$.
\par Another important parameter is the amplitude of curvature perturbation
\begin{equation}
 \Delta^2_R = \frac{8V}{3m_P^4\epsilon}.
\end{equation}

Planck measurement of this parameter is $\Delta^2_R=2.215\times 10^{-9}$ for a wave 
number $k=0.05$ $Mpc^{-1}$. We use this experimental value of  $\Delta^2_R$ to fix  $\lambda^{\prime}$. 

\par
\begin{figure}[h]
\centering{\includegraphics[scale=1.0]{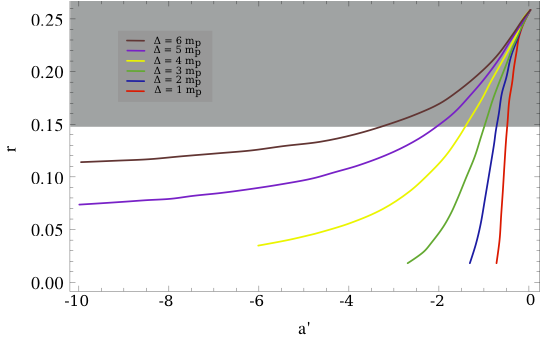}}
\caption{$r$ vs $a^{\prime}$  for several values of  $\Delta$. The region in gray is excluded by Planck}
\label{ra}
\end{figure}

 Let us discuss our results beginning  with  FIG.~\ref{ra}. There  we show the behavior of the scalar to tensor  ratio, $r$, related to $a^{\prime}$ 
for some values of $\Delta$.  First of all, so as to have an idea of the values of $\Phi_i$ and $\Phi_f$, for  the case of $\Delta=3m_P$, and considering the setup presented above, we have  that inflation ends with $\Phi_f\sim 10^{18}$~GeV and, for the particular case of  $60$ e-folds, we get the initial value $\Phi_i\sim 4 \times 10^{19}$~GeV.  Note that as $a^{\prime}$ goes to zero  all the curves converge to a point around $r=0.26$.  This is the expected value for $r$ provided by $\phi^4$ chaotic inflation.  Thus, in our case, the current bounds on $r$ requires $a^{\prime}\neq0$. This means that radiative corrections turns to be absolutely necessary in our analysis.   We also  stress that the scalar to tensor ratio demands trans-planckian regime for $\Delta$ because the sub-Planckian case faces problems in the integration on the e-fold number  to reach the value 60  unless $a^{\prime}$ goes to zero, again recovering the $\Phi^4$  chaotic inflation. Even for the trans-planckian case, on assuming $a^{\prime}<0$,  the current values of $r$ do not allow $\Delta$ to exceed  the regime of $\sim 6m_P$. In other words, our inflation model requires sizable radiative corrections in order to obey the current value of $n_S$ and the bound on $r$. All this run into $a^{\prime}\neq 0 $ and   $\Delta$ around few $m_P$.

\begin{figure}[h]
\centering{\includegraphics[scale=1.1]{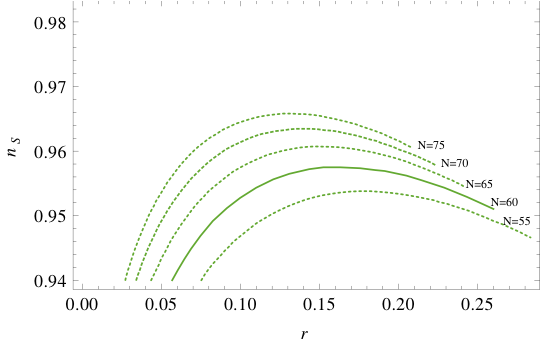}}
\caption{$n_S$ vs $r$  for $\Delta=3 m_P$. }
\label{nsversusr}
\end{figure}

In FIG.~\ref{nsversusr} we present  our results for $n_S$ and $r$ in  a plot confronting $n_s$ with $r$ for $\Delta=3 m_P$ and $a^{\prime}$ obeying the values corresponding to the green curve in FIG.~\ref{ra} for several e-fold values. As we can see in that plot, the model predictions for $n_S$ and $r$ are in perfect  agreement with the experimental bounds  provided by PLANCK2015.  This result is valid for  any other choice of the values for the parameter $\Delta$ presented in FIG.~\ref{ra}.

\begin{figure}[h]
\centering{\includegraphics[scale=0.4]{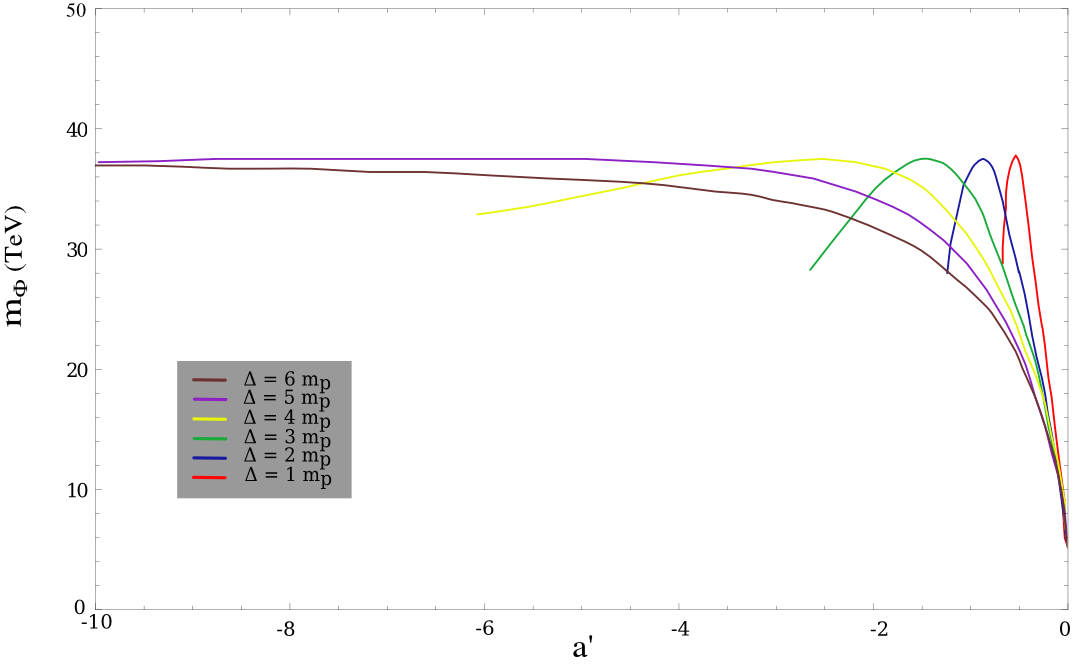}}
\caption{$m_\phi$ vs $a'$   for several values of  $\Delta$. }\label{mphiversusalinha}
\end{figure}

Another interesting outcome we have obtained concerns the inflaton mass. Its expression at tree level is extracted from the diagonalization of the mass matrix $M_R^2$ in Eq.~(\ref{matrixR2}). As reheating demands  very tiny  $\lambda_{\phi}$ and $v_\phi \gg v_\rho \,,\, v_\eta \,,\, v_{\chi^{\prime}}$, then the $(M_R^2)_{44}$ element of that matrix decouples incurring into the following expression for the  inflaton mass at tree level,  $m_{\Phi} \sim \sqrt{2\lambda_{10}}v_\phi$. When radiative corrections are plugged in, this expression receives a correction that depends on the parameters  $a^{\prime}$ and $\Delta$. In FIG.~\ref{mphiversusalinha} we plot the behaviour of the inflaton mass $m_{\Phi}$ with $a^{\prime}$  for some values of $\Delta$.  Even if  $v_\phi$ is around $10^{10}$ GeV, but as the coupling $\lambda_{10}$  is very small, as required by reheating phase, the inflaton   gains a small mass when compared to the conventional chaotic inflation case. According to the prediction of our model, the  inflaton may develop mass until few tens of TeV.  This has implications to the reheating phase, as discussed below. 

\begin{figure}[h]
\centering{\includegraphics[scale=1.0]{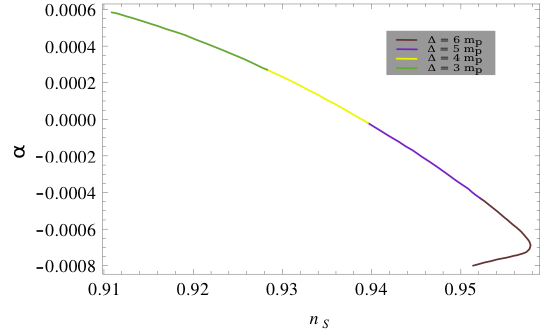}}
\caption{$\alpha$ vs $n_S$  for several values of  $\Delta$.  }\label{alphaversusns}
\end{figure}

For sake of completeness, in FIG.~\ref{alphaversusns} we plot the running index $\alpha$ versus  $n_S$ for some values of $\Delta$. There we have a relatively small  $\alpha$ value 
for all points as it has to be in chaotic inflation. 

We finish this section by discussing reheating~\cite{Abbott:1982hn}.   First of all notice that our inflaton couples to the heavy neutrinos through the Yukawa coupling in Eq.~(\ref{yukintera}), and to scalars through the last four terms in the potential in Eq.~(\ref{vphi1}). Because $v_\phi \sim 10^{10}$~GeV, the inflaton develops mass around tens of TeV, as shown in FIG.~\ref{mphiversusalinha}. This order of magnitude for the inflaton mass forbids that it   decays into a pair of  heavy neutrinos once, as we will see below,  $m_{N_R}\sim 10^7$~GeV. Thus reheating will be solely  due to the decay of the inflaton into a pair of scalars.  

The  3-3-1 model in question involves several scalars, which makes it very difficult to analytically obtain  the scalars in the physical basis. Because of this we  just estimate the reheating temperature that may be achieved in our model.

Even with the inflaton decaying into a  pair of scalars only, it does not face trouble in reheating the universe until temperatures around $10^9$~GeV which is the highest  temperature that does not present the gravitino problem. For our proposal it is just enough to  parameterize  the coupling among the inflaton and a pair of Higgs, provided by those last four terms in the potential in Eq.~(\ref{vphi1}),  by the general form:   $\frac{\lambda}{8} v_\phi \Phi hh$. According to this coupling, we obtain,
\begin{equation}
\Gamma( \Phi \rightarrow hh)\sim \frac{\lambda^2 v^2_\phi}{32 \pi m_{\Phi}}.
\label{widthdecay}
\end{equation}
As it is well known the reheating temperature is estimated to be
\begin{equation}
T_R \sim 0.1\sqrt{\Gamma( \Phi \rightarrow hh) m_P }.
\label{TR}
\end{equation}
For $v_\phi =10^{10}$~GeV and $m_{\Phi}\sim 10$~TeV, a reheating temperature around $10^9$~GeV requires $\lambda \sim 10^{-6}$. This means that the couplings  $ \lambda_{\phi\,,\, 11\,,\,12\,,\,13}$ must must be around this order of magnitude.  Such tiny values for these couplings is typical in chaotic inflation models. In summary, in spite of the fact that the inflaton has an unusual small mass, the model is efficient in reheating the universe.

\section{Some remarks and conclusions}
\label{sec3}
When $\eta^0$ and $\phi$ develop VEVs, the last two terms in the Lagrangian in Eq.~(\ref{yukintera}) yields Dirac and Majorana mass terms for $\nu_L$ and $N_R$,
\be
{\cal L} \supset M_D \bar{\nu}_{L} N_{R} +M \bar{N}^C_{R} N_{R} +H.c.,
\label{masstermsnu}
\ee
where $M_D=h\frac{v_\eta}{\sqrt{2}}$ and $M=\frac{h^{\prime}v_\phi}{\sqrt{2}}$.  These terms provides the following mass matrix for the six massive neutrino, 
\begin{equation}
M_\nu=
\begin{pmatrix}
0 & M_D  \\
M^T_D &  M
\end{pmatrix}.
\label{SSmatrix}
\end{equation}
This is the well known mass matrix for the type I seesaw mechanism whose the diagonalization, for $M \gg M_D$,  leads to~\cite{GellMann:1980vs}\cite{Mohapatra:1979ia},
\be
m_{\nu_L}\simeq \frac{M^2_D}{M}\,\,\,\,\,\,\,\mbox{and}\,\,\,\,\,\\,\,M_R\simeq M. 
\label{seesaw}
\ee
Here we are interested only in getting an estimate on its order of magnitude. As  $a^{\prime}\approx \frac{a}{32\pi^2\lambda^{\prime}}$ and 
$a=-16 h^{\prime 4}$,  for $a^{\prime}$ around $10$ and $\lambda^{\prime} \sim 10^{-14}$, as required by $\Delta^2_R=2.215\times 10^{-9}$,   we get  $h^{\prime} \sim 10^{-3}$, which results in $M_R \sim 10^7$~GeV. So as to obtain heavy neutrinos with such a mass  and standard neutrinos at eV scale, in agreement with solar and atmospheric neutrino oscillation, we just need   $M_D \sim (10^{-1}-  10^{-2})$~GeV. This is obtained for $h$ in the range $\sim (10^{-3 }-10^{-4})$ for $v_\eta \sim 10^2$~GeV. Such range of values for $h$  are of the same order of the average Yukawa couplings in the standard model.

Axion dark matter is considered as an attractive alternative to thermal WIMP dark matter. Our axion is invisible and receives mass through  chiral anomaly,  $m^2_{\mbox{axion}} \sim \frac{\Lambda^4_{QCD}}{f^2_{pq}}$, which gives mass around $10^{-3}$ eV for  $f_{pq} \sim 10^{10}$ GeV and $\Lambda_{QCD} \sim 10^{-1}$ GeV, turning  our axion  a natural candidate for cold dark matter.  As PQ symmetry is broken during inflation, our axion will be produced in the early universe through the misalignment mechanism and its relic abundance is cast  in Refs.~\cite{Turner:1990uz} \cite{Linde:1991km}.

Just few words about  heavy neutrinos with masses around $10^7$~GeV. These neutrinos interact with charged  scalars, as allowed by the Yukawa coupling   $h\bar{f}_{L}\eta N_{R}$, and may give rise to baryogenesis through leptogenesis. Because of the complexity and importance of such subject, we treat it separately elsewhere. However, for a previous treatment of this issue in a similar situation, but different scenario,  we refer the reader to the Ref.~\cite{Huong:2015dwa} .

In summary, several papers have proposed extensions of the standard model that provide  a common origin to the understanding of the strong CP-problem,  dark matter, inflation, and  small neutrino masses. In this paper we argued that  such proposal is elegantly realized in the framework of a  3-3-1 gauge model. In it the strong CP-problem is solved with the PQ symmetry whose associated axion is invisible and may  constitute the dark matter of the universe. Inflation is driven by the real part of the neutral scalar singlet that contains the axion. Successful inflation was obtained by considering radiative corrections to the inflaton potential.  The model predicts an inflaton with  mass of tens of GeV that may be probed in a future  100~TeV proton-proton colision. Reheating is achieved through the decay of the  inflaton into scalars, solely and neutrinos gain small mass through the type I seesaw mechanism. 
\acknowledgments
This work was supported by Conselho Nacional de Pesquisa e
Desenvolvimento Cient\'{i}fico- CNPq (C.A.S.P,  P.S.R.S. ) and Coordena\c c\~ao de Aperfei\c coamento de Pessoal de N\'{i}vel Superior - CAPES (J.G.R.). 

\bibliography{inflation331.bib}

\begin{thebibliography}{39}
\expandafter\ifx\csname natexlab\endcsname\relax\def\natexlab#1{#1}\fi
\expandafter\ifx\csname bibnamefont\endcsname\relax
  \def\bibnamefont#1{#1}\fi
\expandafter\ifx\csname bibfnamefont\endcsname\relax
  \def\bibfnamefont#1{#1}\fi
\expandafter\ifx\csname citenamefont\endcsname\relax
  \def\citenamefont#1{#1}\fi
\expandafter\ifx\csname url\endcsname\relax
  \def\url#1{\texttt{#1}}\fi
\expandafter\ifx\csname urlprefix\endcsname\relax\def\urlprefix{URL }\fi
\providecommand{\bibinfo}[2]{#2}
\providecommand{\eprint}[2][]{\url{#2}}

\bibitem[{\citenamefont{Frampton}(1992)}]{Frampton:1992wt}
\bibinfo{author}{\bibfnamefont{P.~H.} \bibnamefont{Frampton}},
  \bibinfo{journal}{Phys. Rev. Lett.} \textbf{\bibinfo{volume}{69}},
  \bibinfo{pages}{2889} (\bibinfo{year}{1992}).

\bibitem[{\citenamefont{de~Sousa~Pires and
  Ravinez}(1998)}]{deSousaPires:1998jc}
\bibinfo{author}{\bibfnamefont{C.~A.} \bibnamefont{de~Sousa~Pires}}
  \bibnamefont{and} \bibinfo{author}{\bibfnamefont{O.~P.}
  \bibnamefont{Ravinez}}, \bibinfo{journal}{Phys. Rev.}
  \textbf{\bibinfo{volume}{D58}}, \bibinfo{pages}{035008}
  (\bibinfo{year}{1998}), \bibinfo{note}{[Phys. Rev.D58,35008(1998)]},
  \eprint{hep-ph/9803409}.

\bibitem[{\citenamefont{de~Sousa~Pires}(1999)}]{deSousaPires:1999ca}
\bibinfo{author}{\bibfnamefont{C.~A.} \bibnamefont{de~Sousa~Pires}},
  \bibinfo{journal}{Phys. Rev.} \textbf{\bibinfo{volume}{D60}},
  \bibinfo{pages}{075013} (\bibinfo{year}{1999}), \eprint{hep-ph/9902406}.

\bibitem[{\citenamefont{Pal}(1995)}]{Pal:1994ba}
\bibinfo{author}{\bibfnamefont{P.~B.} \bibnamefont{Pal}},
  \bibinfo{journal}{Phys. Rev.} \textbf{\bibinfo{volume}{D52}},
  \bibinfo{pages}{1659} (\bibinfo{year}{1995}), \eprint{hep-ph/9411406}.

\bibitem[{\citenamefont{Weinberg}(1978)}]{Weinberg:1977ma}
\bibinfo{author}{\bibfnamefont{S.}~\bibnamefont{Weinberg}},
  \bibinfo{journal}{Phys. Rev. Lett.} \textbf{\bibinfo{volume}{40}},
  \bibinfo{pages}{223} (\bibinfo{year}{1978}).

\bibitem[{\citenamefont{Wilczek}(1978)}]{Wilczek:1977pj}
\bibinfo{author}{\bibfnamefont{F.}~\bibnamefont{Wilczek}},
  \bibinfo{journal}{Phys. Rev. Lett.} \textbf{\bibinfo{volume}{40}},
  \bibinfo{pages}{279} (\bibinfo{year}{1978}).

\bibitem[{\citenamefont{Kim}(1979)}]{Kim:1979if}
\bibinfo{author}{\bibfnamefont{J.~E.} \bibnamefont{Kim}},
  \bibinfo{journal}{Phys. Rev. Lett.} \textbf{\bibinfo{volume}{43}},
  \bibinfo{pages}{103} (\bibinfo{year}{1979}).

\bibitem[{\citenamefont{Shifman et~al.}(1980)\citenamefont{Shifman, Vainshtein,
  and Zakharov}}]{Shifman:1979if}
\bibinfo{author}{\bibfnamefont{M.~A.} \bibnamefont{Shifman}},
  \bibinfo{author}{\bibfnamefont{A.~I.} \bibnamefont{Vainshtein}},
  \bibnamefont{and} \bibinfo{author}{\bibfnamefont{V.~I.}
  \bibnamefont{Zakharov}}, \bibinfo{journal}{Nucl. Phys.}
  \textbf{\bibinfo{volume}{B166}}, \bibinfo{pages}{493} (\bibinfo{year}{1980}).

\bibitem[{\citenamefont{Dine et~al.}(1981)\citenamefont{Dine, Fischler, and
  Srednicki}}]{Dine:1981rt}
\bibinfo{author}{\bibfnamefont{M.}~\bibnamefont{Dine}},
  \bibinfo{author}{\bibfnamefont{W.}~\bibnamefont{Fischler}}, \bibnamefont{and}
  \bibinfo{author}{\bibfnamefont{M.}~\bibnamefont{Srednicki}},
  \bibinfo{journal}{Phys. Lett.} \textbf{\bibinfo{volume}{B104}},
  \bibinfo{pages}{199} (\bibinfo{year}{1981}).

\bibitem[{\citenamefont{Montero et~al.}(2002)\citenamefont{Montero,
  De~S.~Pires, and Pleitez}}]{Montero:2001ts}
\bibinfo{author}{\bibfnamefont{J.~C.} \bibnamefont{Montero}},
  \bibinfo{author}{\bibfnamefont{C.~A.} \bibnamefont{De~S.~Pires}},
  \bibnamefont{and} \bibinfo{author}{\bibfnamefont{V.}~\bibnamefont{Pleitez}},
  \bibinfo{journal}{Phys. Rev.} \textbf{\bibinfo{volume}{D65}},
  \bibinfo{pages}{095001} (\bibinfo{year}{2002}), \eprint{hep-ph/0112246}.

\bibitem[{\citenamefont{Dias et~al.}(2012)\citenamefont{Dias, de~S.~Pires,
  Rodrigues~da Silva, and Sampieri}}]{Dias:2012xp}
\bibinfo{author}{\bibfnamefont{A.~G.} \bibnamefont{Dias}},
  \bibinfo{author}{\bibfnamefont{C.~A.} \bibnamefont{de~S.~Pires}},
  \bibinfo{author}{\bibfnamefont{P.~S.} \bibnamefont{Rodrigues~da Silva}},
  \bibnamefont{and} \bibinfo{author}{\bibfnamefont{A.}~\bibnamefont{Sampieri}},
  \bibinfo{journal}{Phys. Rev.} \textbf{\bibinfo{volume}{D86}},
  \bibinfo{pages}{035007} (\bibinfo{year}{2012}), \eprint{1206.2590}.

\bibitem[{\citenamefont{Dong and Long}(2008)}]{Dong:2008sw}
\bibinfo{author}{\bibfnamefont{P.~V.} \bibnamefont{Dong}} \bibnamefont{and}
  \bibinfo{author}{\bibfnamefont{H.~N.} \bibnamefont{Long}},
  \bibinfo{journal}{Phys. Rev.} \textbf{\bibinfo{volume}{D77}},
  \bibinfo{pages}{057302} (\bibinfo{year}{2008}), \eprint{0801.4196}.

\bibitem[{\citenamefont{Boucenna et~al.}(2015)\citenamefont{Boucenna, Valle,
  and Vicente}}]{Boucenna:2015zwa}
\bibinfo{author}{\bibfnamefont{S.~M.} \bibnamefont{Boucenna}},
  \bibinfo{author}{\bibfnamefont{J.~W.~F.} \bibnamefont{Valle}},
  \bibnamefont{and} \bibinfo{author}{\bibfnamefont{A.}~\bibnamefont{Vicente}},
  \bibinfo{journal}{Phys. Rev.} \textbf{\bibinfo{volume}{D92}},
  \bibinfo{pages}{053001} (\bibinfo{year}{2015}), \eprint{1502.07546}.

\bibitem[{\citenamefont{de~S.~Pires and Rodrigues~da
  Silva}(2007)}]{deS.Pires:2007gi}
\bibinfo{author}{\bibfnamefont{C.~A.} \bibnamefont{de~S.~Pires}}
  \bibnamefont{and} \bibinfo{author}{\bibfnamefont{P.~S.}
  \bibnamefont{Rodrigues~da Silva}}, \bibinfo{journal}{JCAP}
  \textbf{\bibinfo{volume}{0712}}, \bibinfo{pages}{012} (\bibinfo{year}{2007}),
  \eprint{0710.2104}.

\bibitem[{\citenamefont{Mizukoshi et~al.}(2011)\citenamefont{Mizukoshi,
  de~S.~Pires, Queiroz, and Rodrigues~da Silva}}]{Mizukoshi:2010ky}
\bibinfo{author}{\bibfnamefont{J.~K.} \bibnamefont{Mizukoshi}},
  \bibinfo{author}{\bibfnamefont{C.~A.} \bibnamefont{de~S.~Pires}},
  \bibinfo{author}{\bibfnamefont{F.~S.} \bibnamefont{Queiroz}},
  \bibnamefont{and} \bibinfo{author}{\bibfnamefont{P.~S.}
  \bibnamefont{Rodrigues~da Silva}}, \bibinfo{journal}{Phys. Rev.}
  \textbf{\bibinfo{volume}{D83}}, \bibinfo{pages}{065024}
  (\bibinfo{year}{2011}), \eprint{1010.4097}.

\bibitem[{\citenamefont{Rodrigues~da Silva}(2014)}]{daSilva:2014qba}
\bibinfo{author}{\bibfnamefont{P.~S.} \bibnamefont{Rodrigues~da Silva}}
  (\bibinfo{year}{2014}), \eprint{1412.8633}.

\bibitem[{\citenamefont{Dong et~al.}(2015)\citenamefont{Dong, Kim, Soa, and
  Thuy}}]{Dong:2015rka}
\bibinfo{author}{\bibfnamefont{P.~V.} \bibnamefont{Dong}},
  \bibinfo{author}{\bibfnamefont{C.~S.} \bibnamefont{Kim}},
  \bibinfo{author}{\bibfnamefont{D.~V.} \bibnamefont{Soa}}, \bibnamefont{and}
  \bibinfo{author}{\bibfnamefont{N.~T.} \bibnamefont{Thuy}},
  \bibinfo{journal}{Phys. Rev.} \textbf{\bibinfo{volume}{D91}},
  \bibinfo{pages}{115019} (\bibinfo{year}{2015}), \eprint{1501.04385}.

\bibitem[{\citenamefont{Huong and Long}(2010)}]{Huong:2008ia}
\bibinfo{author}{\bibfnamefont{D.~T.} \bibnamefont{Huong}} \bibnamefont{and}
  \bibinfo{author}{\bibfnamefont{H.~N.} \bibnamefont{Long}},
  \bibinfo{journal}{Phys. Atom. Nucl.} \textbf{\bibinfo{volume}{73}},
  \bibinfo{pages}{791} (\bibinfo{year}{2010}), \eprint{0807.2346}.

\bibitem[{\citenamefont{Long}(2015)}]{Long:2015qza}
\bibinfo{author}{\bibfnamefont{H.~N.} \bibnamefont{Long}}
  (\bibinfo{year}{2015}), \eprint{1501.01852}.

\bibitem[{\citenamefont{Dias et~al.}(2014)\citenamefont{Dias, Machado, Nishi,
  Ringwald, and Vaudrevange}}]{Dias:2014osa}
\bibinfo{author}{\bibfnamefont{A.~G.} \bibnamefont{Dias}},
  \bibinfo{author}{\bibfnamefont{A.~C.~B.} \bibnamefont{Machado}},
  \bibinfo{author}{\bibfnamefont{C.~C.} \bibnamefont{Nishi}},
  \bibinfo{author}{\bibfnamefont{A.}~\bibnamefont{Ringwald}}, \bibnamefont{and}
  \bibinfo{author}{\bibfnamefont{P.}~\bibnamefont{Vaudrevange}},
  \bibinfo{journal}{JHEP} \textbf{\bibinfo{volume}{06}}, \bibinfo{pages}{037}
  (\bibinfo{year}{2014}), \eprint{1403.5760}.

\bibitem[{\citenamefont{Barenboim and Park}(2016)}]{Barenboim:2015cqa}
\bibinfo{author}{\bibfnamefont{G.}~\bibnamefont{Barenboim}} \bibnamefont{and}
  \bibinfo{author}{\bibfnamefont{W.-I.} \bibnamefont{Park}},
  \bibinfo{journal}{Phys. Lett.} \textbf{\bibinfo{volume}{B756}},
  \bibinfo{pages}{317} (\bibinfo{year}{2016}), \eprint{1508.00011}.

\bibitem[{\citenamefont{Ballesteros
  et~al.}(2016{\natexlab{a}})\citenamefont{Ballesteros, Redondo, Ringwald, and
  Tamarit}}]{Ballesteros:2016euj}
\bibinfo{author}{\bibfnamefont{G.}~\bibnamefont{Ballesteros}},
  \bibinfo{author}{\bibfnamefont{J.}~\bibnamefont{Redondo}},
  \bibinfo{author}{\bibfnamefont{A.}~\bibnamefont{Ringwald}}, \bibnamefont{and}
  \bibinfo{author}{\bibfnamefont{C.}~\bibnamefont{Tamarit}}
  (\bibinfo{year}{2016}{\natexlab{a}}), \eprint{1608.05414}.

\bibitem[{\citenamefont{Ballesteros
  et~al.}(2016{\natexlab{b}})\citenamefont{Ballesteros, Redondo, Ringwald, and
  Tamarit}}]{Ballesteros:2016xej}
\bibinfo{author}{\bibfnamefont{G.}~\bibnamefont{Ballesteros}},
  \bibinfo{author}{\bibfnamefont{J.}~\bibnamefont{Redondo}},
  \bibinfo{author}{\bibfnamefont{A.}~\bibnamefont{Ringwald}}, \bibnamefont{and}
  \bibinfo{author}{\bibfnamefont{C.}~\bibnamefont{Tamarit}}
  (\bibinfo{year}{2016}{\natexlab{b}}), \eprint{1610.01639}.

\bibitem[{\citenamefont{Ade et~al.}(2016)}]{Ade:2015lrj}
\bibinfo{author}{\bibfnamefont{P.~A.~R.} \bibnamefont{Ade}}
  \bibnamefont{et~al.} (\bibinfo{collaboration}{Planck}),
  \bibinfo{journal}{Astron. Astrophys.} \textbf{\bibinfo{volume}{594}},
  \bibinfo{pages}{A20} (\bibinfo{year}{2016}), \eprint{1502.02114}.

\bibitem[{\citenamefont{Dias et~al.}(2003)\citenamefont{Dias, de~S.~Pires, and
  Rodrigues~da Silva}}]{Dias:2003iq}
\bibinfo{author}{\bibfnamefont{A.~G.} \bibnamefont{Dias}},
  \bibinfo{author}{\bibfnamefont{C.~A.} \bibnamefont{de~S.~Pires}},
  \bibnamefont{and} \bibinfo{author}{\bibfnamefont{P.~S.}
  \bibnamefont{Rodrigues~da Silva}}, \bibinfo{journal}{Phys. Rev.}
  \textbf{\bibinfo{volume}{D68}}, \bibinfo{pages}{115009}
  (\bibinfo{year}{2003}), \eprint{hep-ph/0309058}.

\bibitem[{\citenamefont{Singer et~al.}(1980)\citenamefont{Singer, Valle, and
  Schechter}}]{Singer:1980sw}
\bibinfo{author}{\bibfnamefont{M.}~\bibnamefont{Singer}},
  \bibinfo{author}{\bibfnamefont{J.~W.~F.} \bibnamefont{Valle}},
  \bibnamefont{and}
  \bibinfo{author}{\bibfnamefont{J.}~\bibnamefont{Schechter}},
  \bibinfo{journal}{Phys. Rev.} \textbf{\bibinfo{volume}{D22}},
  \bibinfo{pages}{738} (\bibinfo{year}{1980}).

\bibitem[{\citenamefont{Montero et~al.}(1993)\citenamefont{Montero, Pisano, and
  Pleitez}}]{Montero:1992jk}
\bibinfo{author}{\bibfnamefont{J.~C.} \bibnamefont{Montero}},
  \bibinfo{author}{\bibfnamefont{F.}~\bibnamefont{Pisano}}, \bibnamefont{and}
  \bibinfo{author}{\bibfnamefont{V.}~\bibnamefont{Pleitez}},
  \bibinfo{journal}{Phys. Rev.} \textbf{\bibinfo{volume}{D47}},
  \bibinfo{pages}{2918} (\bibinfo{year}{1993}), \eprint{hep-ph/9212271}.

\bibitem[{\citenamefont{Foot et~al.}(1994)\citenamefont{Foot, Long, and
  Tran}}]{Foot:1994ym}
\bibinfo{author}{\bibfnamefont{R.}~\bibnamefont{Foot}},
  \bibinfo{author}{\bibfnamefont{H.~N.} \bibnamefont{Long}}, \bibnamefont{and}
  \bibinfo{author}{\bibfnamefont{T.~A.} \bibnamefont{Tran}},
  \bibinfo{journal}{Phys. Rev.} \textbf{\bibinfo{volume}{D50}},
  \bibinfo{pages}{R34} (\bibinfo{year}{1994}), \eprint{hep-ph/9402243}.

\bibitem[{\citenamefont{Coleman and Weinberg}(1973)}]{Coleman:1973jx}
\bibinfo{author}{\bibfnamefont{S.~R.} \bibnamefont{Coleman}} \bibnamefont{and}
  \bibinfo{author}{\bibfnamefont{E.~J.} \bibnamefont{Weinberg}},
  \bibinfo{journal}{Phys. Rev.} \textbf{\bibinfo{volume}{D7}},
  \bibinfo{pages}{1888} (\bibinfo{year}{1973}).

\bibitem[{\citenamefont{Senoguz and Shafi}(2008)}]{NeferSenoguz:2008nn}
\bibinfo{author}{\bibfnamefont{V.~N.} \bibnamefont{Senoguz}} \bibnamefont{and}
  \bibinfo{author}{\bibfnamefont{Q.}~\bibnamefont{Shafi}},
  \bibinfo{journal}{Phys. Lett.} \textbf{\bibinfo{volume}{B668}},
  \bibinfo{pages}{6} (\bibinfo{year}{2008}), \eprint{0806.2798}.

\bibitem[{\citenamefont{Boucenna et~al.}(2014)\citenamefont{Boucenna, Morisi,
  Shafi, and Valle}}]{Boucenna:2014uma}
\bibinfo{author}{\bibfnamefont{S.~M.} \bibnamefont{Boucenna}},
  \bibinfo{author}{\bibfnamefont{S.}~\bibnamefont{Morisi}},
  \bibinfo{author}{\bibfnamefont{Q.}~\bibnamefont{Shafi}}, \bibnamefont{and}
  \bibinfo{author}{\bibfnamefont{J.~W.~F.} \bibnamefont{Valle}},
  \bibinfo{journal}{Phys. Rev.} \textbf{\bibinfo{volume}{D90}},
  \bibinfo{pages}{055023} (\bibinfo{year}{2014}), \eprint{1404.3198}.

\bibitem[{\citenamefont{Liddle et~al.}(1994)\citenamefont{Liddle, Parsons, and
  Barrow}}]{Liddle:1994dx}
\bibinfo{author}{\bibfnamefont{A.~R.} \bibnamefont{Liddle}},
  \bibinfo{author}{\bibfnamefont{P.}~\bibnamefont{Parsons}}, \bibnamefont{and}
  \bibinfo{author}{\bibfnamefont{J.~D.} \bibnamefont{Barrow}},
  \bibinfo{journal}{Phys. Rev.} \textbf{\bibinfo{volume}{D50}},
  \bibinfo{pages}{7222} (\bibinfo{year}{1994}), \eprint{astro-ph/9408015}.

\bibitem[{\citenamefont{Lyth and Liddle}(2009)}]{Lyth:2009zz}
\bibinfo{author}{\bibfnamefont{D.~H.} \bibnamefont{Lyth}} \bibnamefont{and}
  \bibinfo{author}{\bibfnamefont{A.~R.} \bibnamefont{Liddle}},
  \emph{\bibinfo{title}{{The primordial density perturbation: Cosmology,
  inflation and the origin of structure}}} (\bibinfo{year}{2009}),
  \urlprefix\url{http://www.cambridge.org/uk/catalogue/catalogue.asp?isbn=9780521828499}.

\bibitem[{\citenamefont{Abbott et~al.}(1982)\citenamefont{Abbott, Farhi, and
  Wise}}]{Abbott:1982hn}
\bibinfo{author}{\bibfnamefont{L.~F.} \bibnamefont{Abbott}},
  \bibinfo{author}{\bibfnamefont{E.}~\bibnamefont{Farhi}}, \bibnamefont{and}
  \bibinfo{author}{\bibfnamefont{M.~B.} \bibnamefont{Wise}},
  \bibinfo{journal}{Phys. Lett.} \textbf{\bibinfo{volume}{B117}},
  \bibinfo{pages}{29} (\bibinfo{year}{1982}).

\bibitem[{\citenamefont{Gell-Mann et~al.}(1979)\citenamefont{Gell-Mann, Ramond,
  and Slansky}}]{GellMann:1980vs}
\bibinfo{author}{\bibfnamefont{M.}~\bibnamefont{Gell-Mann}},
  \bibinfo{author}{\bibfnamefont{P.}~\bibnamefont{Ramond}}, \bibnamefont{and}
  \bibinfo{author}{\bibfnamefont{R.}~\bibnamefont{Slansky}},
  \bibinfo{journal}{Conf. Proc.} \textbf{\bibinfo{volume}{C790927}},
  \bibinfo{pages}{315} (\bibinfo{year}{1979}), \eprint{1306.4669}.

\bibitem[{\citenamefont{Mohapatra and Senjanovic}(1980)}]{Mohapatra:1979ia}
\bibinfo{author}{\bibfnamefont{R.~N.} \bibnamefont{Mohapatra}}
  \bibnamefont{and}
  \bibinfo{author}{\bibfnamefont{G.}~\bibnamefont{Senjanovic}},
  \bibinfo{journal}{Phys. Rev. Lett.} \textbf{\bibinfo{volume}{44}},
  \bibinfo{pages}{912} (\bibinfo{year}{1980}).

\bibitem[{\citenamefont{Turner and Wilczek}(1991)}]{Turner:1990uz}
\bibinfo{author}{\bibfnamefont{M.~S.} \bibnamefont{Turner}} \bibnamefont{and}
  \bibinfo{author}{\bibfnamefont{F.}~\bibnamefont{Wilczek}},
  \bibinfo{journal}{Phys. Rev. Lett.} \textbf{\bibinfo{volume}{66}},
  \bibinfo{pages}{5} (\bibinfo{year}{1991}).

\bibitem[{\citenamefont{Linde}(1991)}]{Linde:1991km}
\bibinfo{author}{\bibfnamefont{A.~D.} \bibnamefont{Linde}},
  \bibinfo{journal}{Phys. Lett.} \textbf{\bibinfo{volume}{B259}},
  \bibinfo{pages}{38} (\bibinfo{year}{1991}).

\bibitem[{\citenamefont{Huong et~al.}(2015)\citenamefont{Huong, Dong, Kim, and
  Thuy}}]{Huong:2015dwa}
\bibinfo{author}{\bibfnamefont{D.~T.} \bibnamefont{Huong}},
  \bibinfo{author}{\bibfnamefont{P.~V.} \bibnamefont{Dong}},
  \bibinfo{author}{\bibfnamefont{C.~S.} \bibnamefont{Kim}}, \bibnamefont{and}
  \bibinfo{author}{\bibfnamefont{N.~T.} \bibnamefont{Thuy}},
  \bibinfo{journal}{Phys. Rev.} \textbf{\bibinfo{volume}{D91}},
  \bibinfo{pages}{055023} (\bibinfo{year}{2015}), \eprint{1501.00543}.

\end{thebibliography}
\end{document}